\documentclass[11pt,a4paper]{article}

\usepackage[margin=1in]{geometry}
\usepackage{amsmath}
\usepackage{newtxtext}
\usepackage{newtxmath}
\usepackage{microtype}
\usepackage{caption}
\usepackage{url}
\usepackage[numbers,sort&compress]{natbib}
\usepackage{authblk}

\usepackage{verbatim}
\usepackage{graphicx}
\usepackage{booktabs}
\usepackage{multirow}
\usepackage[dvipsnames,svgnames]{xcolor}
\colorlet{punct}{red!60!black}
\definecolor{background}{HTML}{EEEEEE}
\definecolor{delim}{RGB}{20,105,176}
\colorlet{numb}{magenta!60!black}
\graphicspath{{./Figures/}{./}}
\usepackage{listings} 
\lstdefinelanguage{json}{
    basicstyle=\small\ttfamily,
    numbers=left,
    numberstyle=\scriptsize,
    stepnumber=1,
    numbersep=8pt,
    showstringspaces=false,
    breaklines=true,
    frame=lines,
    backgroundcolor=\color{background},
    literate=
     *{0}{{{\color{numb}0}}}{1}
      {1}{{{\color{numb}1}}}{1}
      {2}{{{\color{numb}2}}}{1}
      {3}{{{\color{numb}3}}}{1}
      {4}{{{\color{numb}4}}}{1}
      {5}{{{\color{numb}5}}}{1}
      {6}{{{\color{numb}6}}}{1}
      {7}{{{\color{numb}7}}}{1}
      {8}{{{\color{numb}8}}}{1}
      {9}{{{\color{numb}9}}}{1}
      {:}{{{\color{punct}{:}}}}{1}
      {,}{{{\color{punct}{,}}}}{1}
      {\{}{{{\color{delim}{\{}}}}{1}
      {\}}{{{\color{delim}{\}}}}}{1}
      {[}{{{\color{delim}{[}}}}{1}
      {]}{{{\color{delim}{]}}}}{1},
}

\usepackage{tikz}
\newcommand{\tikzxmark}{%
\tikz[scale=0.23] {
    \draw[line width=0.7,line cap=round] (0,0) to [bend left=6] (1,1);
    \draw[line width=0.7,line cap=round] (0.2,0.95) to [bend right=3] (0.8,0.05);
}}
\usepackage{makecell}
\usepackage{vcell}

\usepackage{siunitx}
\usepackage{booktabs}

\title{Federated Detection of Open Charge Point Protocol 1.6 Cyberattacks}

\author[1,2]{Christos Dalamagkas}
\author[3]{Panagiotis Radoglou-Grammatikis\thanks{Corresponding author: pradoglou@uowm.gr; ORCID: 0000-0003-1605-9413.}}
\author[4]{Pavlos Bouzinis}
\author[2]{Ioannis Papadopoulos}
\author[1]{Thomas Lagkas}
\author[5]{Vasileios Argyriou}
\author[6]{Sotirios Goudos}
\author[7]{Dimitrios Margounakis}
\author[7]{Eleftherios Fountoukidis}
\author[3]{Panagiotis Sarigiannidis}
\affil[1]{Department of Computer Science, Democritus University of Thrace, 65404 Kavala, Greece}
\affil[2]{Innovation Hub, Public Power Corporation S.A., 15351 Kantza-Pallini, Greece}
\affil[3]{Department of Electrical and Computer Engineering, University of Western Macedonia, 50150 Kozani, Greece}
\affil[4]{Metamind Innovations IKE, 50150 Kozani, Greece}
\affil[5]{Department of Networks and Digital Media, Kingston University London, Surrey KT1 2EE, UK}
\affil[6]{School of Physics, Aristotle University of Thessaloniki, 54124 Thessaloniki, Greece}
\affil[7]{SIDROCO Holdings Ltd, 1082 Nicosia, Cyprus}

\date{}

\usepackage{hyperref}
\hypersetup{colorlinks=true, allcolors=blue}

\begin{document}
\maketitle

\begin{abstract}
The ongoing electrification of the transportation sector requires the deployment of multiple Electric Vehicle (EV) charging stations across multiple locations. However, the EV charging stations introduce significant cyber-physical and privacy risks, given the presence of vulnerable communication protocols, like the Open Charge Point Protocol (OCPP). Meanwhile, the Federated Learning (FL) paradigm showcases a novel approach for improved intrusion detection results that utilize multiple sources of Internet of Things data, while respecting the confidentiality of private information. This paper proposes  an FL-based intrusion detection system, which leverages OCPP 1.6 network flows to detect OCPP 1.6 cyberattacks. The evaluation results showcase high detection performance of the proposed FL-based solution. 


\end{abstract}

\section{Introduction}
\label{sec:Introduction}

Electric Vehicles (EVs) are a driving force towards the electrification of the transportation sector, contributing to the reduction of the environmental footprint and towards achieving the sustainability goals. In this context, the European Union (EU) plans to ban new non-electric cars starting from the year 2035. At the same time, the deployment of EV Charging Stations (EVCSs) increases in order to ensure seamless experience for EV users as well as to reduce the range anxiety \cite{Rauh2014}. The EV charging is associated with numerous services, including management of the EVCSs, handling and billing of charging transactions, metering, roaming of EV charging services as well as communication with the power grid and the Distribution/Transmission System Operators (DSOs/TSOs). 

To realize properly integrated and interoperable EV charging services, a number of protocols and standards are in force. For example, International Standards Organization (ISO) / International Electro-technical Commission (IEC) 15118 is proposed for defining the Vehicle-to-Grid interface for bi-directional charging/discharging of EVs, smart charging and plug \& charge. Roaming between Charging Station Operators (CSO) and e-mobility service providers is accomplished by the Open Charge Point Interface (OCPI) standard, whereas grid operators can interact with the CSOs through demand response protocols, e.g., the Open Automated Demand Response (openADR). 

While many of the above mentioned standards and protocols are optional or their development is on-going, a fundamental interaction within the EV charging ecosystem is the one between EVCSs and the EV Charging Station Management System (EVCSMS) (operated by the CSO), through the Open Charge Point Protocol (OCPP). OCPP is an open standard, maintained by the Open Charge Alliance (OCA), for the vendor-neutral remote management and monitoring of EVCSs. Common operations of OCPP include the authorization and management of transactions as well as the maintenance of the EVCSs (e.g., firmware upgrades and system logs monitoring). Currently, the version 1.6 of OCPP is the most widely deployed version of the protocol, it is supported by the majority of EVCS and EVCSMS manufacturers as well as the one that is fully certified by the OCA.

Despite its significance, OCPP is associated with notable cybersecurity concerns. For example, C. Alcaraz et al. \cite{Alcaraz2017} assess the security features of OCPP, potential vulnerabilities and threat scenarios resulting from the protocol design and characteristics. The authors investigate the implementation and feasibility of various threats, including False Data Injections (FDI), Man-in-The-Middle (MiTM), impersonation, data tampering, fraud / energy theft, and Denial of Service (DoS), by developing a virtual infrastructure based on multiple Virtual Machines (VMs) to replicate the EVCSs and the OCPP server. For the identified threats, the authors explain how they could be materialized as well as corresponding mitigation measures. The authors extend this analysis for the newest version of the standard, OCPP 2.0.1, in \cite{Alcaraz2023}.  
A detailed security assessment of the EV charging ecosystem is performed by A. Kaur et. al \cite{Kaur2025}. The authors discuss the attack vector and threat models for multiple protocols used in the EV charging domain, including ISO-15118, OCPP, OCPI and openADR. For OCPP, six relevant threat models are identified. In particular, a MiTM is an imminent threat, in which a potential attacker can be placed between the EVCSMS and the EVCS in order to intercept unencrypted OCPP traffic. A DoS attack is also feasible by introducing a fuzzer, as highlighted in OCPPStorm \cite{Coppoletta2024}, by injecting malformed or unexpected inputs in the OCPP communication, causing unexpected behavior, system dysfunction, crashes and operational instability. A botnet attack is also feasible, by utilizing multiple compromised EVCS that target the EVCSMS or other assets of the EV charging network through malware spreading. Finally, through impersonation, an attacker could intercept and replay sensitive data that is used for authenticating EVCS and authorizing charging transactions. In this way, a malicious actor could impersonate a legitimate EVCS by using its unique ID or initiate charging transaction by using the user ID of another legitimate user.

Considering the critical security issues with respect to OCPP, an open challenge remains the dynamic detection of potential threats and cyberattacks. In this paper, we focus our attention on detecting potential cyberattacks against OCPP 1.6, by adopting the Federated Learning (FL) approach for analyzing OCPP-based network flows. FL is a distributed Artificial Intelligence (AI) system, in which multiple entities train their local AI model and contribute to the training of a global AI model \cite{Zhang2022}. Compared to the conventional approach of a single and central AI model, the FL paradigm is characterized by enhanced performance as well as for respecting data privacy and data access rights. Considering the private and sensitive data that could be transferred or be elicited by analyzing OCPP traffic (e.g., EV user identity, charging locations, behavioral patterns), the FL architecture is a suitable solution for privacy-aware security analysis and threat detection, benefiting from the contribution of multiple clients realized as multiple EV charging hubs.

Based on the aforementioned remarks, the contribution of this paper is summarized as follows:

\begin{itemize}
    \item We describe four OCPP cyberattacks and provide insights with respect to their implementation and observations that can be leveraged for their detection.
    \item We develop the \texttt{OCPPFlowMeter}, a flow-based network analysis tool that generates network flows with OCPP 1.6 features, enabling the detection of attacks covering not only the network and transmission layers but also the OCPP 1.6 application layer. 
    \item We propose an FL-based IDS architecture that can be applied to monitor the OCPP 1.6 traffic of multiple EVCSs, grouped as EV charging hubs at multiple locations. The evaluation results showcase high detection accuracy, ranging from $98.49\%$ to $99.21\%$, obtained by comparing the results of six FL aggregation methods.
    \item As a result of this work, we have developed and published the \textit{Federated OCPP 1.6 Intrusion Detection Dataset} \cite{federatedocppDataset}, which can also be found on Zenodo\footnote{\url{https://zenodo.org/records/14887131}}. This dataset includes network traffic samples as well as the respective flow statistics of the four OCPP 1.6 cyberattacks developed in this paper.
\end{itemize}

The rest of this paper is organized as follows. Section 1.1 discusses the existing works on FL-based intrusion detection  and detection of OCPP threats, Section 2 presents the proposed adoption of FL on the EV charging infrastructure as well as the OCPP cyberattacks that we consider in this work, and the proposed FL-based IDS. Section 3 presents the evaluation results,  section 4 discusses the results and potential next steps for future research directions and improvements, and section 5 concludes this work by providing the key takeaways.

\subsection{Related Work}
\label{sec:Related Work}
This section discusses the related work with respect to a) FL adoption in the context of cybersecurity, and b) existing methods for detecting OCPP cyberattacks.

\subsubsection{FL for cybersecurity}

V. Mothukuri et al. \cite{Mothukuri2022} propose a privacy-focused FL approach for Internet of Things (IoT). Each IoT device trains a Gate Recurrent Unit (GRU) model using its local data, and sends the parameters to a central FL server that aggregates them and sends the updated weights to each IoT device. As a result, the accuracy of each model increases, while the data of each IoT device remains private. The authors evaluate the performance of their solution by using a public dataset based on Modbus TCP, which is processed by CICFlowMeter to extract the features that the AI models can use to classify the attacks. According to the evaluation results,  the proposed FL approach achieves 90.2\% accuracy, showcasing 4.1\% improvement on accuracy compared to the non FL approach.

In \cite{Rashid2023}, M. Rashid el at. introduce a dynamic weighted aggregation federated learning (DAFL), a new aggregation technique that implements dynamic filtering and weighting strategies for local models. The proposed method improves the performance of conventional FL-based IDS in terms of communication overhead, while maintaining high detection accuracy and preserving data privacy.

In \cite{Idrissi2023}, M. Idrissi et al. propose the Federated Anomaly-Based Network Intrusion Detection System (Fed-ANIDS), which employs auto-encoders and FL in a distributed manner. The proposed architecture further employs two aggregation methods, FedProx and FedAvg. The proposed system is evaluated using numerous public datasets, namely the \texttt{USTC-TFC2016}, \texttt{CIC-IDS2017}, and \texttt{CSE-CIC-IDS2018} datasets, which use an updated version of \texttt{CICFlowMeter} that improves the construction of network flows. The proposed solution is also compared with Generative Adversarial Network (GAN) models, and the results indicated a better detection performance, in terms of higher accuracy and fewer false alarms. In addition, the results showcased that FedProx delivered better results than FedAvg.

A. Karunamurthy et. al \cite{Karunamurthy2025} introduce a deep learning FL-based IDS for IoT environments. In the proposed architecture, the local models in the remote IoT environments leverage Convolutional Neural Networks (CNNs) with different parameter settings. Furthermore, an optimal feature selection model resides on the federated aggregation server, based on the Chimp optimization method. Inspired by the behavior of chimps in their natural environment, the feature selection model applies a fitness function to select the most optimal features, which are given as input to the deep learning classifier. The proposed model is evaluated on an MQTT protocol dataset, which includes 5 different attacks. The proposed method achieves 93.3\% recall, 94\% F1 score and 95.5\% accuracy. 

Finally, P. Radoglou et. al \cite{RadoglouGrammatikis2025} present a multimodal FL-based IDS, called AI4FIDS, which is able to analyze four different data types, namely system logs, operational data, network flow statistics, and visual representations, in order to detect potential anomalies or indication of cyberattacks in the critical infrastructure. Combining the multiple detection methods, AI4FIDS includes a majority voting and a weighted majority method aggregate the predictions of the multiple underlying models in order to generate a single, combined prediction. 

\subsubsection{Detection of OCPP cyberattacks}
Moreover, several works have been identified, which describe OCPP threats and try to address their detection. In more detail, A. Morosan et al. \cite{Morosan2017} propose a Back-Propagation Neural Network (BPNN) that is able to classify EVCSs between the normal and faulted states, based on the OCPP 1.5-J traffic they generate. The three-layered neural network was able to determine a faulty EVCS based a) on the similarity of consecutive pairs of request-response, and b) the OCPP message type from the server. In a similar approach in \cite{Kabir2021}, Kabir et al. propose a BPNN utilized by the EVCSMS to analyze the OCPP requests and detect potential malicious attempts of coordinated switching attacks, i.e. charging/discharging and back again within a very short time period.

Mansi et al. \cite{Mansi2022} aim to predict and mitigate cyberattacks against EVCSs, therefore, they propose a cybersecurity framework that predicts and mitigates potential cyberattacks. In particular, the authors employ the STRIDE thread modeling to predict potential vulnerabilities and attack vectors on EVCSs, then a weighted attack defense tree is developed to analyze the adversary's objectives and create attack scenarios. A Hidden Markov Model is proposed in order to predict the possible attack paths, while a Partially Observable Monte-Carlo Planning (POMCP) algorithm ensures that the attacker is directed towards the predicted paths, ensuring that mitigation actions are timely placed to reduce the impact of the attack. While the proposed method can predict the attacker's behavior in multi-step attack scenarios, the problem on how to classify the individual behavior as malicious is not addressed.

M. Elkashlan et al. \cite{ElKashlan2023} utilize the \texttt{IoT-23} dataset in order to demonstrate the detection of cyberattacks against EVCSs. The authors employ various machine learning algorithms to detect malicious traffic that could be associated with threats against EVCSs, including Command and Control server communication, Distributed Denial of Service attack, traces of the Okiru botnet, and samples of a horizontal port scan. The authors discussed the features of the dataset and removed those that presented weak correlation, hence choosing 14 out of 21 features. The authors get results in terms of Precision, Accuracy, Recall and F-1 scores, by comparing four classifiers: The Naive Bayes, the J48 classifier, the attribute-select classifier, and the filtered classifier. However, the proposed methodology did not address EVCS-specific attacks, while the authors highlight the necessity of building a dedicated dataset for EVCSs.

The \texttt{DataTransfer} operation of OCPP is utilized by some works in order to implement custom security logic, extending the capabilities of OCPP 1.6. In particular, Rubio et al. \cite{Rubio2018} aim to ensure the confidentiality and integrity of the energy values exchanged during charging transactions, by using \texttt{DataTransfer} messages to enable the EVCSMS to exchange secrets with the EVCS. After exchanging secrets, different cryptographic mechanisms can be applied to securely exchange energy values through \texttt{meterValue} and \texttt{meterStop} messages. D. Kim et. al \cite{Kim2025} also utilize \texttt{DataTransfer} to provide security services, in particular, for implementing a rollback mechanism on the EVCSs, thus mitigating the impact of potential threats such as replay attacks, maliciously modified requests and unauthorized activities.

I. Benfarhat et. al \cite{Benfarhat2025} leverage deep learning and propose a temporal convolutional network (TCN) in order to classify up to 17 different cyberattacks against the ISO-15118 and OCPP protocols, as they are introduced in the CICEVSE2024 dataset \cite{Buedi2024}. According to the authors, the proposed model provides benefits in terms of representing temporal dependencies and spatial characteristics, thus being able to detect complex patterns. The authors evaluate their solution by measuring accuracy, precision, recall and F1-score metrics, while comparing the results with CNN, DNN, RNN and LSTM models as well as a hybrid CNN-DNN-LSTM model. Three experiments are conducted, while considering two classes, five classes and 17 classes respectively. In the binary and 5-class classification experiments, all metrics reach 100\%. On the contrary, when considering 17 classes, the TCN achieves 93\% in all metrics. Despite the large number of classes considered by the authors, it is noteworthy that the attacks considered exhibit common attack vectors of the network and transport layer, such as network scanning with the nmap tool and TCP/UDP flooding variants, which are already addressed by other solutions considering IoT-based DDoS attack detection \cite{Bala2024}.

M. Rahman et. al \cite{Rahman2024} evaluate a hybrid architecture of CNN and LSTM networks to detect the attacks of the CICEVSE2024 dataset. To enhance the detection performance as well as to increase the understanding of the dataset and attack patterns, the authors apply the SHAP explainability method to obtain results on the features that contributed mostly on the detection outcomes. It is worth mentioning that the selected features are related solely to transport-layer characteristics, such as the average packet inter-arrival time in network flows. The evaluation results indicate high performance, achieving accuracy, precision, recall and F1 score at 97.5\% and Area Under Curve (AUC) at 98.5\%.

Finally, S. Purohit et. al \cite{Purohit2024} propose an FL-based IDS for detecting cyberattacks against the EVCS infrastructure. Similar to \cite{Benfarhat2025} and \cite{Rahman2024}, the authors utilize the CICEVSE2024 dataset in order to evaluate the performance of their solution, achieving an accuracy of around 97\% and superior F1-score compared to centralized AI-based IDS that use the CICEVSE2024 dataset.

Summarizing, multiple works have been identified that aim to develop AI-based solutions for detecting attacks in IoT and EVCS setups. However, it is noteworthy that most of the existing work\cite{ElKashlan2023, Benfarhat2025, Buedi2024, Rahman2024, Purohit2024} employ a similar methodology of detecting OCPP attacks by analyzing network flow features stemming from the network and transport layers of the TCP/IP stack, for example those generated by CICFlowMeter\cite{Engelen2021} and NFStream\footnote{\url{https://www.nfstream.org/}}. However, this approach is not sufficient for detecting threats that rely on application-layer characteristics, while the network and transport layers remain agnostic. Moreover, the threat model considered by the aforementioned works and the CICEVSE2024 dataset focus mostly on common IoT threats rather than on OCPP-specific attack scenarios.

\begin{figure}[htbp]
    \centering
    \includegraphics[width=0.8\textwidth,page=1]{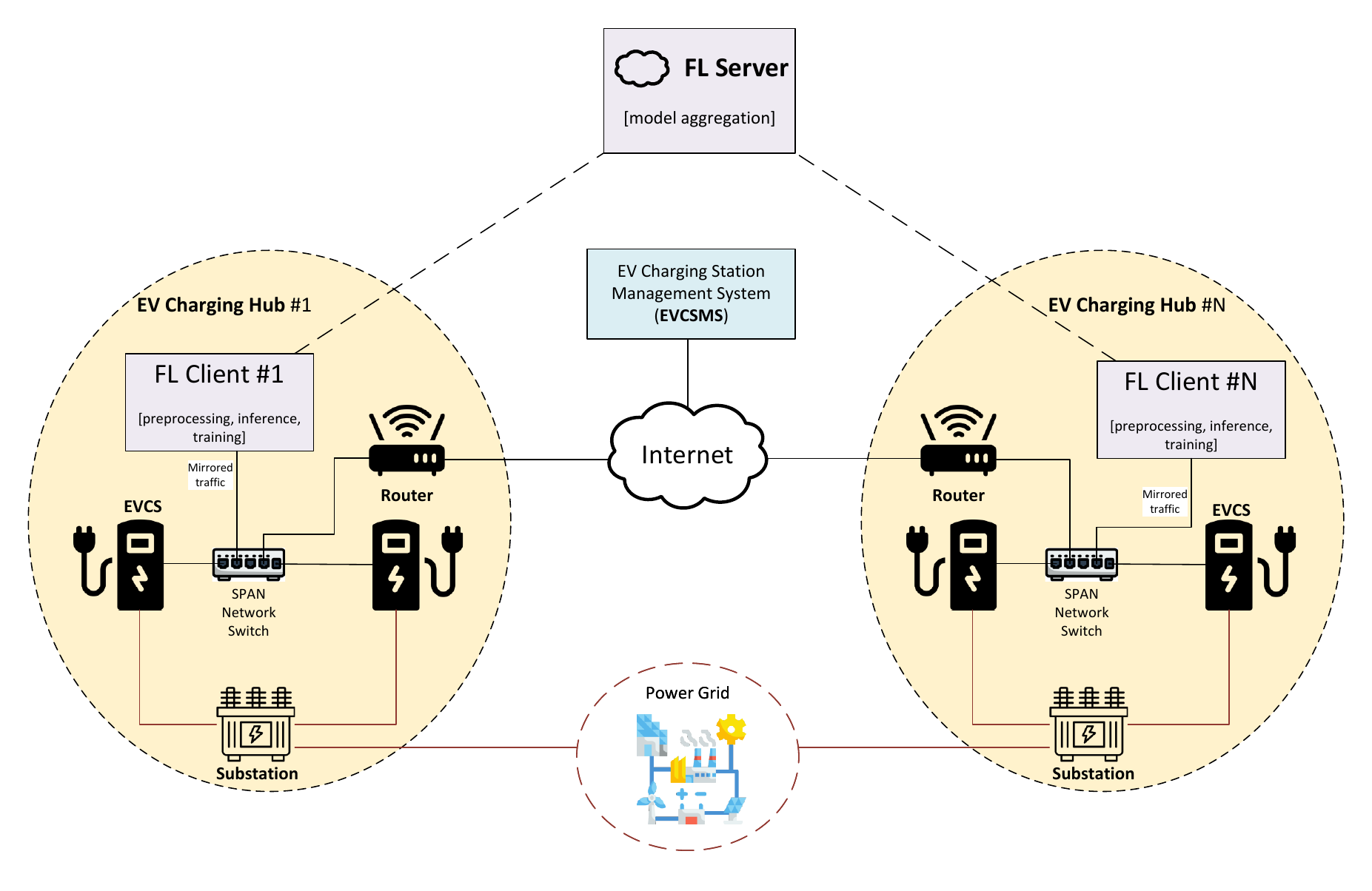}
    \caption[fl-architecture]{The FL-based IDS architecture applied on the EV charging infrastructure. Used icons from: \url{https://www.flaticon.com/}}
    \label{fig:fl-architecture}
\end{figure}

\section{Methods}
\subsection{Case Study}
\label{sec:Case Study}
In this work, we introduce a novel approach for detecting OCPP cyberattacks, by applying the FL architecture on the EV charging infrastructure  and detecting potential threats by generating and analyzing network flows with OCPP-related features. \autoref{fig:fl-architecture} depicts the proposed solution applied on the EV charging infrastructure. The system under study consists of multiple locations, where multiple EV charging stations are deployed and serve EV users. EVCSs in the same location form an EV charging hub, which could correspond to a shopping mall or an airport parking area. Each EV charging hub is connected to the interconnected power grid through a distribution substation. In each EV charging hub, Internet connectivity is provided by a gateway router, necessary for interconnecting the EVCSs with the EVCSMS. Owned by the CSO, the EVCSMS controls the EVCSs through the OCPP 1.6 protocol. Finally, on each EV charging hub, an FL client is deployed, which is connected to the FL server. The FL client receives the raw network traffic from the EV charging hub, containing OCPP traffic traces of the EVCSs, and analyses the traffic for potential cyberattacks. The local models residing on the FL clients are updated by the centralized FL server, by considering the updates coming from all the EV charging hubs.

\subsection{OCPP 1.6 Cyberattacks}
\label{sec:OCPP Cyberattacks}
\begin{figure}[htbp]
    \centering
    \includegraphics[width=0.8\textwidth,page=3]{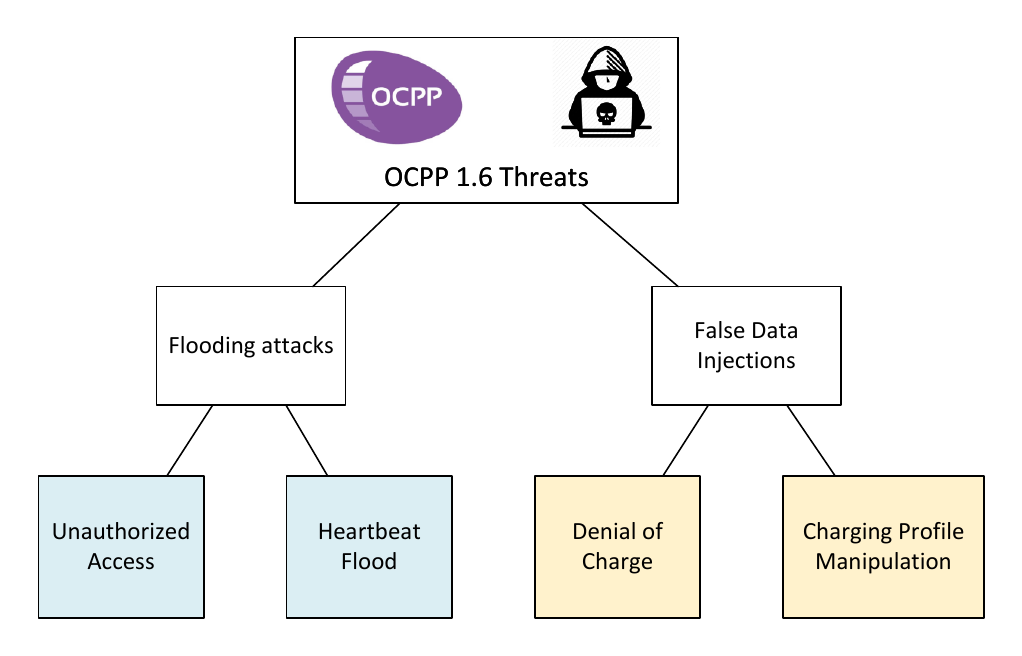}
    \caption[ocpp-cyberattacks]{The OCPP cyberattacks considered in this work}
    \label{fig:ocpp-cyberattacks}
\end{figure}

According to the works relevant to OCPP threats mentioned in Section 1, we have considered the implementation of four cyberattacks grouped into two categories, namely Flooding attacks and False Data Injection (FDI) attacks. Under the FDI category, we consider: a) Charging Profile Manipulation, and b) Denial of Charge, while for the Flooding attacks, we consider: c) Unauthorized Access, and d) Heartbeat Flood. These attacks are summarized and illustrated in ~\autoref{fig:ocpp-cyberattacks}.

The main difference between the two categories is that the flooding attacks rely on overwhelming the target with application-layer network packets, which the target is unable to properly handle. Depending on possible implementation flaws or weaknesses from the side of the target, these attacks may cause exhaustion of computing resources and unavailability of services. On the contrary, the FDIs depend on the fact that the OCPP 1.6 packets are transmitted unencrypted and unsigned. Based on this fact, a malicious insider places themself between the EVCSMS and the EVCSs through Address Resolution Protocol (ARP) cache poisoning \cite{Kelli2022} in order to alter the OCPP 1.6 messages being transmitted. The consequences vary, depending on the modification carried out by the adversary. The FDIs considered in this work cause either denial of service or lead to cyber-physical consequences. 

The cyberattacks are described in more detail in the next subsections. For each cyberattack, we describe: 1) the \textit{normal operation}, i.e., the respective flow of operations based on the OCPP 1.6 standard, 2) the \textit{threat(s)} that we identify based on potential flaws or weaknesses of the normal operation, 3) the \textit{cyberattack}, which describes how we materialize the threat, and 4) \textit{observations} of the cyberattack in terms of traces or abnormal behavior that could be considered for detecting the cyberattack. 

\subsubsection{Charging Profile Manipulation}
\textbf{Normal operation}: The \texttt{SetChargingProfile.req} message is an OCPP 1.6 operation that is used by the EVCSMS to install charging profiles to EVCSs. A charging profile describes the amount of power or current that an EVCS is allowed to deliver per time interval. According to the OCPP 1.6 specification, this operation can be issued either in the context of a charging transaction (i.e., at the start or during the transaction) or outside the context of a transaction, as a separate message \cite{ocpp16}. The \texttt{SetChargingProfile.req} message includes a \texttt{csChargingProfiles} object, that defines the charging schedule. Multiple time intervals are defined as separate items inside the \texttt{chargingSchedulePeriod} list. The example provided in \autoref{fig:SetChargingProfile.req} describes a \texttt{SetChargingProfile.req} that installs a charging profile, which instructs connector 1 of an EVCS to draw at most $15A$ on 2024-05-12, from 13:51:54 to 15:51:54. 
 
\begin{figure}[htbp]
    \caption{Example of a SetChargingProfile.req message}
    \label{fig:SetChargingProfile.req}
    \begin{lstlisting}[language=json,frame=single,numbers=none]
{"connectorId": 1,
"csChargingProfiles": {
    "chargingProfileId": 1,
    "transactionId": null,
    "stackLevel": 1,
    "chargingProfilePurpose": "TxDefaultProfile",
    "chargingProfileKind": "Absolute",
    "validFrom":"2024-05-12T13:51:54.037000Z",
    "validTo": "2024-05-12T15:51:54.037000Z",
    chargingSchedule": {
        "duration": 86400,
        "schedulingUnit": "A",
        "chargingSchedulePeriod": {
            "startPeriod": 0,
            "limit": 15,
            "numberPhases": 3
        }
    }
}}
    \end{lstlisting}
\end{figure}

As a powerful and flexible operation, \texttt{SetChargingProfile.req} is used to implement smart charging scenarios and apply complex charging patterns to EVCSs. Moreover, it can also be leveraged by system operators (i.e., DSOs and TSOs), in combination with openADR \cite{ocpp-and-openadr}, for ancillary services, e.g., to reduce peak demands or to avoid a predicted load surge \cite{Sarieddine2023}.

\textbf{Threat}: Given its criticality for the above mentioned reasons, an erroneous or maliciously altered charging profile could have significant cyber-physical impact to the power grid. For example, a CSO may have introduced charging profiles as a safety measure to avoid overloading and stressing of legacy electrical infrastructure, including old electric cables or unmaintained transformers. In that case, false charging profiles may lead to stressing of the electrical infrastructure and potential malfunctions. More sophisticated attacks are also possible, e.g., a coordinated oscillatory load attack that could manipulate the load following an on/off pattern, causing system frequency fluctuations that can threaten the grid stability \cite{Sarieddine2023}.

\textbf{Cyberattack}: The Charging Profile Manipulation attack we consider in this work assumes that a cyberattacker performs MiTM, through ARP poisoning, followed by FDI that modifies the \texttt{SetChargingProfile.req} messages being transmitted from the EVCSMS to the EVCSs. For each incoming \texttt{SetChargingProfile.req} message, the attacker replaces the value of the limit attribute of all \texttt{chargingSchedulePeriod} objects with a higher number. As a result, the affected EVCS is able to draw more power than originally configured by the CSO.

\textbf{Observation}: The Charging Profile Manipulation FDI could be detected by inspecting the \texttt{limit} values of the transmitted charging profiles. Based on the characteristics of the EV charging infrastructure and the targeted EVCS, there are reasonable values that are expected to be set in this field. By profiling this attribute and getting the baseline from benign traffic, it would be possible for an CSO to detect an abnormal charging profile.

\subsubsection{Denial of Charge}
\begin{figure}[htbp]
    \centering
    \includegraphics[width=0.75\textwidth]{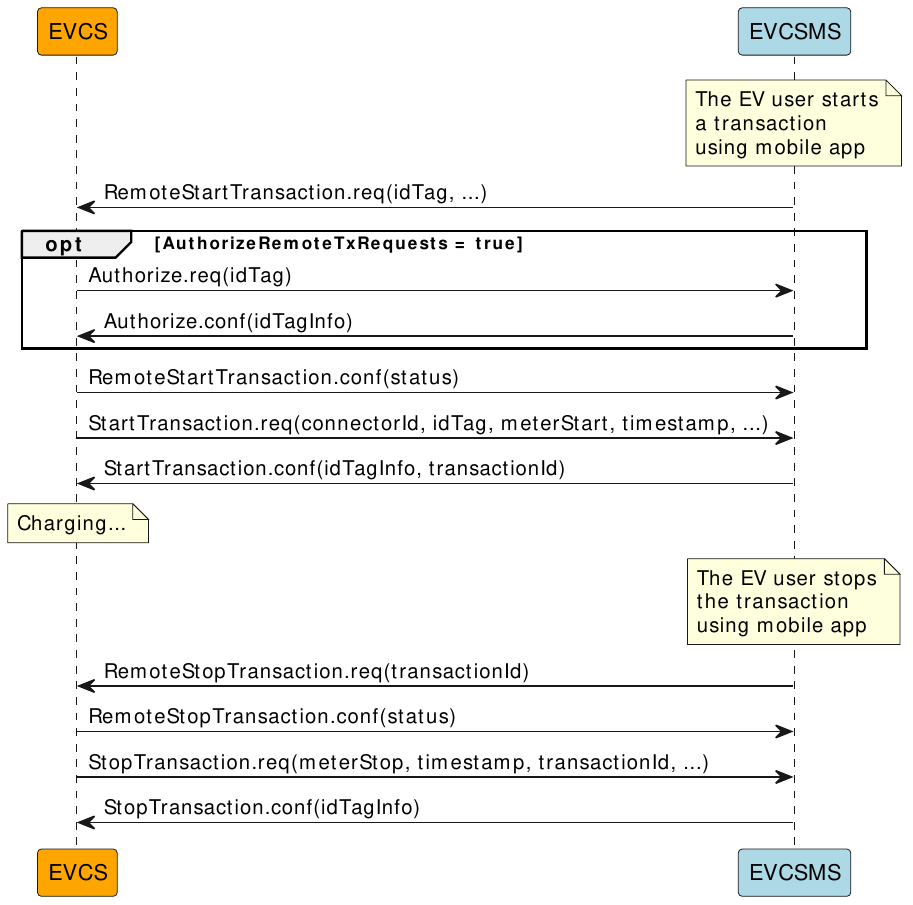}
    \caption[ocpp-cyberattacks]{The authorization of charging transactions in OCPP 1.6}
    \label{fig:doc-flow}
\end{figure}

\textbf{Normal operation}: \autoref{fig:doc-flow} describes the procedure to remotely start a transaction as well as the authorization process before starting a charging session. To start a charging transaction, the EV user through the mobile app will trigger EVCSMS to send a \texttt{RemoteStartTransaction.req} message. If the \texttt{AuthorizeRemoteTxRequests} configuration variable is activated on the EVCS, the EVCS will try to authorize the identity of the EV user (\texttt{idTag}) via an \texttt{Authorize.req} message. Assuming that the presented identity is valid, the EVCS will receive a positive \texttt{Authorize.conf} response and will start the transaction procedure by sending a \texttt{StartTransaction.req} message. The EVCS will start providing power upon receiving a \texttt{StartTransaction.conf} with \texttt{idTagInfo.status = Accepted} from the EVCSMS. Similarly, the transaction can be stopped by the EV user by triggering a \texttt{RemoteStopTransaction.req} message \cite{ocpp16}.

\textbf{Threat}: By observing the authorization procedure, a denial of service (in particular, a denial of charge) situation could happen if the authentication information, which is private information, is tampered with. In particular, if the \texttt{idTag} information is altered during transmission, this will lead to failure of authorization, therefore, preventing the EVCS from starting the charging transaction. 

\textbf{Cyberattack}: We consider a denial of charge attack, in which a cyberattacker performs MiTM, through ARP poisoning, followed by FDI that replaces the \texttt{idTag} info included in any \texttt{RemoteStartTransaction} message with a random value. If \texttt{AuthorizeRemoteTxRequests} on the EVCS is enabled, the EVCS will send an \texttt{Authorize.req} message with the \texttt{idTag} injected by the attacker, leading to an \texttt{Authorize.conf} response with \texttt{idTagInfo.status = Invalid}. If \texttt{AuthorizeRemoteTxRequests} is not enabled on the EVCS, the authorization will still fail, since the EVCSMS  will send a \texttt{StartTransaction.conf} with \texttt{idTagInfo.status = Invalid}.

\textbf{Observation}: Considering that the CSO prefers to minimize the messages transmitted from/to the EVCS, it is reasonable to assume that, normally, an EVCSMS will never send a \texttt{RemoteStartTransaction.req} with an invalid \texttt{idTag}, since the EVCSMS has already the capacity to validate an \texttt{idTag} in the first place. Hence, observing a \texttt{RemoteStartTransaction.req} followed by a \texttt{StartTransaction.conf} or an \texttt{Authorize.conf} message with \texttt{idTagInfo.status = Invalid}, would be an indication of a malformed \texttt{RemoteStartTransaction.req}.

\subsubsection{Heartbeat Flood}
\label{sec:Heartbeat Flood}

\begin{figure}[htbp]
    \centering
    \includegraphics[width=0.75\textwidth]{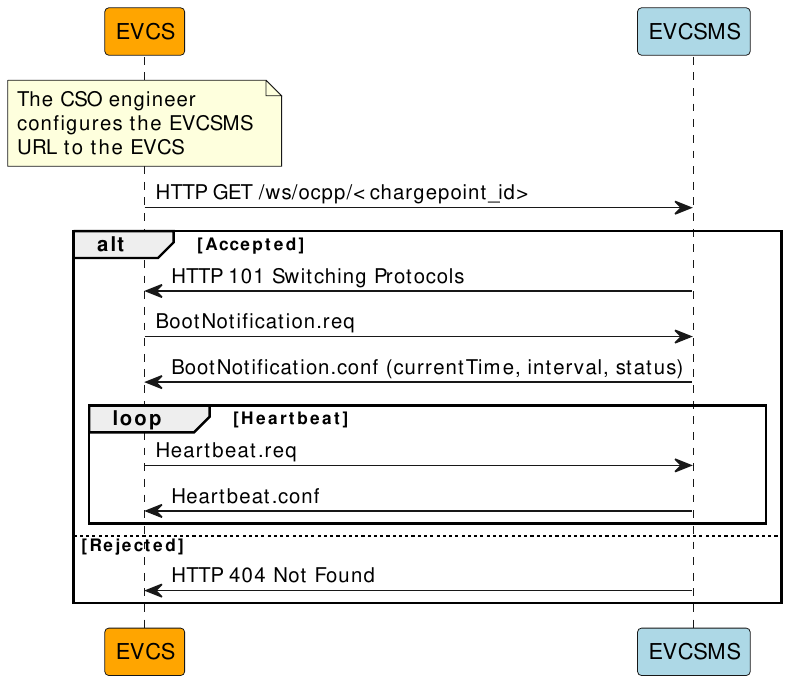}
    \caption[ocpp-cyberattacks]{The establishment of an OCPP 1.6-J session over WebSocket.}
    \label{fig:websocket-session-flow}
\end{figure}

\textbf{Normal operation}: \autoref{fig:websocket-session-flow} depicts an overview of the procedure for establishing an OCPP 1.6 session over WebSocket (OCPP 1.6-J) between an EVCS and the EVCSMS \cite{ocpp16j}. The procedure is initiated by the CSO engineer, which configures the EVCS through an out-of-band management method, usually via Bluetooth and a dedicated mobile app provided by the EVCS manufacturer. The engineer specifies the Unique Resource Identifier (URL) that the EVCS must use to register to the EVCSMS. Upon applying this configuration, the EVCS sends an HyperText Transport Protocol (HTTP) GET request, incorporating the unique ID of the EVCS at the end of the URL as well as the appropriate HTTP headers that request session upgrade to WebSocket. If the EVCSMS accepts the EVCSMS, it will send an HTTP 101 Switching Protocols response, which signals the EVCS to immediately start sending OCPP 1.6-J messages, starting with the \texttt{BootNotification.req}. Depending on the value of the \texttt{HeartbeatInterval} configuration variable of the EVCS and the \texttt{interval} parameter of the \texttt{BootNotification.conf} response of the EVCSMS (which can override the \texttt{HeartbeatInterval}), the EVCS sends periodic \texttt{Heartbeat.req} messages to the EVCSMS in order to get the current date and time from the EVCSMS. Moreover, this message is useful for the EVCSMS to ensure that the EVCS is still online.

\textbf{Threat}: While the Heartbeat interval can be indicated by the EVCSMS via the \texttt{BootNotification.conf} response, right after the OCPP 1.6-J session establishment, a malicious or misconfigured EVCS might not respect this interval, thus sending very frequent Heartbeat messages. Alternatively, in a MiTM situation, an attacker could modify the \texttt{interval} attribute of the \texttt{BootNotification.conf} messages, thus forcing the EVCSs to send very frequent Heartbeat messages. Regardless of the method, the high volume of Hearbeat messages may threaten the availability of the EVCSMS.

\textbf{Cyberattack}: In this attack scenario, we assume that the attacker deploys a botnet of multiple virtual EVCSs, which concurrently attempt to establish OCPP 1.6-J sessions with the targeted EVCSMS. Assuming that the EVCSMS is not configured to authorize each EVCS or that the bots use valid IDs, all connections are accepted. After this step, the bots flood the EVCSMS with \texttt{Heartbeat.req} messages, leading to resource exhaustion and service unavailability for the EVCSMS. For example, such flooding could lead to reaching maximum database connections internally in the EVCSMS, causing unavailability of other critical services provided by the EVCSMS. Moreover, the computing resources could be overutilized (including Central Processing Unit (CPU) cycles and network bandwidth), leading to overall system performance degradation.

\textbf{Observation}: In contrast to the FDI attacks, someone could notice the traces of this attack also at the Transmission Control Protocol (TCP) / Internet Protocol (IP) layer. In particular, an unusually high number of packets will be noticed per TCP session. Moreover, at the application layer, the CSO would notice an unusually high number of Heartbeat messages per EVCS.

\subsubsection{Unauthorized Access}
\textbf{Normal operation}: As an alternative flow depicted in \autoref{fig:websocket-session-flow}, and according to the OCPP 1.6-J specification \cite{ocpp16j}, if an EVCS session establishment attempt is not accepted, then the EVCSMS should reply with an "HTTP 404 - Not Found" response. 

\textbf{Threat}: If the EVCSMS does not apply any throttling policy, an overwhelming amount of connection attempts may cause exhaustion of computing resources and degradation of system performance. Moreover, an attacker could perform an enumeration attack by trying to "guess" valid EVCS IDs.

\textbf{Cyberattack}: Similarly to the Heartbeat Flood attack, in this attack scenario we assume that the attacker deploys a botnet of multiple virtual EVCSs, which concurrently attempt to establish OCPP 1.6-J sessions with the targeted EVCSMS. However, in this variation of the attack, the targeted EVCSMS is appropriately configured in order to reject connection attempts from unknown EVCSs. Hence, the EVCS responds to each bot with an HTTP 404 message, leading to wastage of computing resources and possible performance degradation due to over-utilization of computing and network resources.

\textbf{Observation}: At the TCP/IP layer, this attack would generate an unusually high number of short-lived TCP sessions finished by TCP packets having the FIN flag activated. Moreover, at the application layer, this attack would generate an unusually high number of HTTP 404 messages, clearly indicating failed connection attempts from WebSocket clients. 

\subsection{Federated Learning Intrusion Detection System}
\label{sec:Federated Learning Intrusion Detection System}

\begin{figure}[htbp]
    \centering
    \includegraphics[width=0.8\textwidth,page=2]{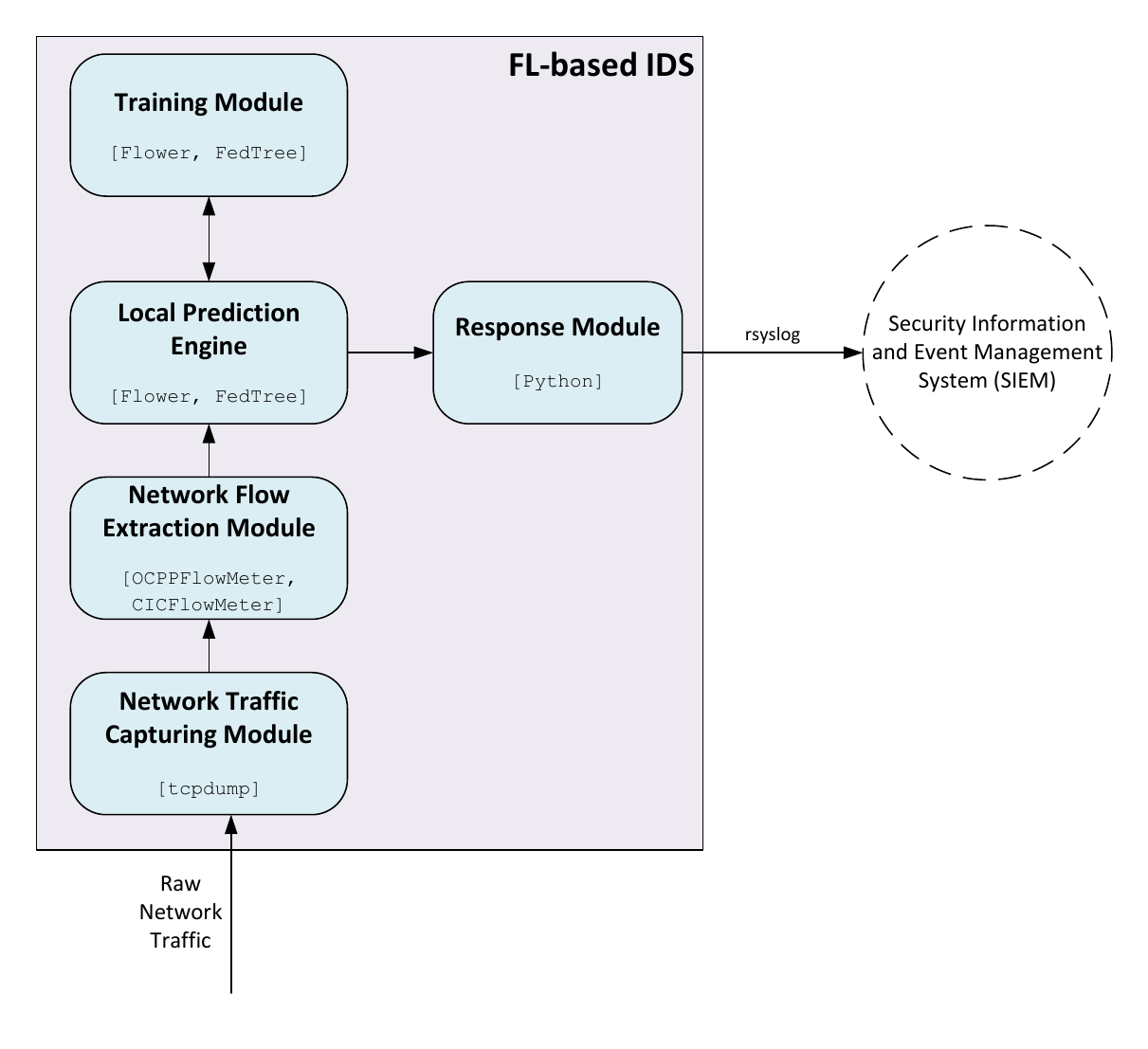}
    \caption{The architecture of the FL-based IDS}
    \label{fig:fl-ids}
\end{figure}

\autoref{fig:fl-ids} depicts the architecture and implementation details of the proposed FL-based IDS, which is composed of the following components: a) the Network Traffic Capturing Module, b) the Network Flow Extraction Module, c) the Local Prediction Engine, and d) the Response Module. In summary, the FL client receives network traffic from the EV charging infrastructure and generates a security event for each abnormal network flow. Finally, the security events are delivered to the preferred Security Information and Event Management (SIEM) system.

\subsubsection{Network Traffic Capturing Module}
\label{sec:Network Traffic Capturing Module}
First, the Network Traffic Capturing Module is responsible for capturing the network traffic data, using a Switched Port Analyser (SPAN) (i.e., port mirroring) of the local network switch. SPAN allows the FL-based IDS to monitor and capture the traffic data passing through specific ports of the network switch, where the EVCSs are connected. In the context of SPAN, the monitoring source and destination ports should be defined. Next, all incoming and outgoing traffic data from the monitoring sources are copied/mirrored to the destination port. Next, the Network Traffic Capturing Module uses \texttt{tcpdump} in order to capture the mirrored network traffic data.

\subsubsection{Network Flow Extraction Module}
\label{sec:Network Flow Extraction Module}
The Network Flow Extraction Module is composed of a set of flow statistics/features generators that receive the network traffic data (i.e., PCAP file) from the previous module and produces bi-directional flow statistics/features. For this purpose, we use two flow generators, namely a) the \texttt{CICFlowMeter}\cite{Engelen2021} tool, b) the \texttt{OCPPFlowMeter}. 

\textbf{\texttt{OCPPFlowMeter}}: It is a custom tool, introduced in this work, that generates additional flow statistics focusing on the OCPP 1.6 protocol characteristics. The complete list of the \texttt{OCPPFlowMeter} features can be found in \autoref{tab:Features of the OCPPFlowMeter}. Compared to \texttt{CICFlowMeter}, which provides statistics only for the IP and TCP network layers, \texttt{OCPPFlowMeter} can effectively be used to detect both flooding and FDI attacks. Some of the features of the \texttt{OCPPFlowMeter} are influenced by the cyberattack observation described for each cyberattack in section 2.2. Adopting the \texttt{CICFlowMeter} approach, the \texttt{OCPPFlowMeter} groups packets based on a pre-defined flow timeout of 120 seconds, and calculates statistics by parsing the payload of the OCPP 1.6 messages as well as the WebSocket and HTTP headers.

\subsubsection{Local Prediction Engine}
\label{sec:Local Prediction Engine}
 Upon completing the FL process, each participating client obtains a local copy of the final global model from the server, during the final FL round. This model is then independently deployed on the client’s infrastructure (e.g., devices, on-premise servers, virtual machines, or cloud-based systems) to perform real-time inference. As such,
the Local Prediction Engine consists of a set of federated detection models that are generated by the Training Module. As input, it receives the flow statistics/features from the previous module and applies the corresponding federated detection models. The Local Prediction Engine model is trained with TCP/IP flow statistics/features from \texttt{CICFlowMeter} and OCPP 1.6 statistics from the \texttt{OCPPFlowMeter}. In case of a detected cyberattack, the Response Module generates the corresponding security event.

\subsubsection{Response Module}
\label{sec:Response Module}
Based on the detection results, the Response Module is responsible for generating the corresponding security events. The security event is delivered to the configured SIEM using the \texttt{rsyslog} protocol.

\subsubsection{Training Module}
\label{sec:Training Module}
The Training Module consists of two components: a) Federated Server (Fed Server) and b) Federated Client(s) (Fed Clients). On the one hand, the Federated Server is responsible for coordinating the FL process and managing the communication between the Federated Clients. It aggregates the locally trained models from the Federated Clients and updates the global model. It also handles the management of resources and data privacy. On the other hand, a Federated Client is responsible for training the AI models on the local data and communicating the trained models to the Federated Server. It also handles the pre-processing and post-processing of the data and the management of the local resources. It is placed on the EV charging hubs to enable the use of the local data for training the models and to ensure the privacy and security of the EV charging data. For the implementation of the Training Module, \texttt{Flower} and \texttt{FedTree} were utilized. Moreover, various aggregation techniques were investigated, such as FedAvg \cite{Zhou2022}, FedProx \cite{li2020federated}, FedAdam \cite{reddi2021adaptive}, FedAdagrad \cite{reddi2021adaptive}, FedYogi \cite{reddi2021adaptive} and FedTree \cite{fedtree}.  Although a detailed descriptions of these techniques are available in their respective references, we outline their main characteristics below. FedAvg is the standard FL aggregation strategy. It works by averaging the model weights from the clients after each local training round and updating the global model accordingly. FedProx extends FedAvg by adding a proximal term in the local objective to limit the deviation of local models from the global model. It is suitable for handling non-independent and identically distributed (non-iid) data among clients. Moreover, FedAdam applies adaptive learning rates by using first and second moment estimates of gradients, FedAdagrad adjusts learning rates based on historical information, and FedYogi maintains stable model updates by controlling the growth of second-moment estimates. These methods focus on providing training stability. Finally, FedTree is designed for decision tree-based models. 

\begin{table}[htbp]
\centering
\caption{Features of the \texttt{OCPPFlowMeter}}
\label{tab:Features of the OCPPFlowMeter}
\resizebox{\columnwidth}{!}{%
\begin{tabular}{@{}clll@{}}
\toprule
\multicolumn{1}{l}{\textbf{Network Layer}} & \textbf{\#} & \textbf{Feature}                                                & \textbf{Description}                                                                                                                \\ \midrule
\multirow{20}{*}{TCP/IP}            & 1  & \texttt{flow\_id}                                               & Unique ID of the flow                                                                                                      \\ \cmidrule(l){2-4} 
                                    & 2  & \texttt{src\_ip}                                                & Source IP of the flow                                                                                                      \\ \cmidrule(l){2-4} 
                                    & 3  & \texttt{dst\_ip}                                                & Destination IP of the flow                                                                                                 \\ \cmidrule(l){2-4} 
                                    & 4  & \texttt{src\_port}                                              & Source TCP port                                                                                                            \\ \cmidrule(l){2-4} 
                                    & 5  & \texttt{dst\_port}                                              & Destination TCP port                                                                                                       \\ \cmidrule(l){2-4} 
                                    & 6  & \texttt{total\_flow\_packets}                                   & Total number of packets contained within the flow                                                                          \\ \cmidrule(l){2-4} 
                                    & 7  & \texttt{total\_fw\_packets}                                     & Total flow packets in the forward direction                                                                                \\ \cmidrule(l){2-4} 
                                    & 8  & \texttt{total\_bw\_packets}                                     & Total flow packets in the backward direction                                                                               \\ \cmidrule(l){2-4} 
                                    & 9  & \texttt{flow\_duration}                                         & Flow duration in seconds                                                                                                   \\ \cmidrule(l){2-4} 
                                    & 10 & \texttt{flow\_down\_up\_ratio}                                  & The fraction between the packets in the backward direction and the packets in the   forward direction                      \\ \cmidrule(l){2-4} 
                                    & 11 & \texttt{flow\_total\_SYN\_flag}                                 & The total number of the TCP SYN packets                                                                                    \\ \cmidrule(l){2-4} 
                                    & 12 & \texttt{flow\_total\_RST\_flag}                                 & The total number of the TCP RST packets                                                                                    \\ \cmidrule(l){2-4} 
                                    & 13 & \texttt{flow\_total\_PSH\_flag}                                 & The total number of the TCP PSH packets                                                                                    \\ \cmidrule(l){2-4} 
                                    & 14 & \texttt{flow\_total\_ACK\_flag}                                 & The total number of the TCP ACK packets                                                                                    \\ \cmidrule(l){2-4} 
                                    & 15 & \texttt{flow\_total\_URG\_flag}                                 & The total number of the TCP URG packets                                                                                    \\ \cmidrule(l){2-4} 
                                    & 16 & \texttt{flow\_total\_CWE\_flag}                                 & The total number of the TCP CWE packets                                                                                    \\ \cmidrule(l){2-4} 
                                    & 17 & \texttt{flow\_total\_ECE\_flag}                                 & The total number of the TCP ECE packets                                                                                    \\ \cmidrule(l){2-4} 
                                    & 18 & \texttt{flow\_total\_FIN\_flag}                                 & The total number of the TCP FIN packets                                                                                    \\ \cmidrule(l){2-4} 
                                    & 19 & \texttt{flow\_start\_timestamp}                                 & The timestamp of the flow. It is defined with the first relevant packet.                                                   \\ \cmidrule(l){2-4} 
                                    & 20 & \texttt{flow\_end\_timestamp}                                   & The timestamp of the last packet of the flow                                                                               \\ \midrule
\multirow{4}{*}{HTTP}               & 21 & \texttt{flow\_total\_http\_get\_packets}                        & The total number of HTTP GET packets                                                                                       \\ \cmidrule(l){2-4} 
                                    & 22 & \texttt{flow\_total\_http\_2xx\_packets}                        & The total number of HTTP 2XX success messages                                                                              \\ \cmidrule(l){2-4} 
                                    & 23 & \texttt{flow\_total\_http\_4xx\_packets}                        & The total number of HTTP 4XX client error messages                                                                         \\ \cmidrule(l){2-4} 
                                    & 24 & \texttt{flow\_total\_http\_5xx\_packets}                        & The total number of HTTP 5XX server error messages                                                                         \\ \midrule
\multirow{10}{*}{WebSocket}         & 25 & \texttt{flow\_websocket\_packts\_per\_second}                   & The number of WebSocket packets per second                                                                                 \\ \cmidrule(l){2-4} 
                                    & 26 & \texttt{fw\_websocket\_packts\_per\_second}                     & The number of WebSocket packets per second in the forward direction                                                        \\ \cmidrule(l){2-4} 
                                    & 27 & \texttt{bw\_websocket\_packts\_per\_second}                     & The number of WebSocket packets per second in the backward direction                                                       \\ \cmidrule(l){2-4} 
                                    & 28 & \texttt{flow\_websocket\_bytes\_per\_second}                    & The sum of WebSocket payload lengths per second                                                                            \\ \cmidrule(l){2-4} 
                                    & 29 & \texttt{fw\_websocket\_bytes\_per\_second}                      & The sum of WebSocket payload lengths per second in the forward direction                                                   \\ \cmidrule(l){2-4} 
                                    & 30 & \texttt{bw\_websocket\_bytes\_per\_second}                      & The sum of WebSocket payload lengths per second in the backward direction                                                  \\ \cmidrule(l){2-4} 
                                    & 31 & \texttt{flow\_total\_websocket\_ping\_packets}                  & The total number of the WebSocket ping packets (opcode 0x9)                                                                \\ \cmidrule(l){2-4} 
                                    & 32 & \texttt{flow\_total\_websocket\_pong\_packets}                  & The total number of the WebSocket pong packets (opcode 0xA)                                                                \\ \cmidrule(l){2-4} 
                                    & 33 & \texttt{flow\_total\_websocket\_close\_packets}                 & The total number of the WebSocket close packets (opcode 0x8)                                                               \\ \cmidrule(l){2-4} 
                                    & 34 & \texttt{flow\_total\_websocket\_data\_messages}                 & The total number of the WebSocket data frames (opcode 0x1 or 0x2)                                                          \\ \midrule
\multirow{20}{*}{OCPP 1.6}          & 35 & \texttt{flow\_total\_ocpp16\_heartbeat\_packets}                & The total number of the OCPP 1.6 Heartbeat messages                                                                        \\ \cmidrule(l){2-4} 
                                    & 36 & \texttt{flow\_total\_ocpp16\_resetHard\_packets}                & The total number of the OCPP 1.6 HardReset messages                                                                        \\ \cmidrule(l){2-4} 
                                    & 37 & \texttt{flow\_total\_ocpp16\_resetSoft\_packets}                & The total number of the OCPP 1.6 SoftReset messages                                                                        \\ \cmidrule(l){2-4} 
                                    & 38 & \texttt{flow\_total\_ocpp16\_unlockconnector\_packets}          & The total number of the OCPP 1.6 UnlockConnector messages                                                                  \\ \cmidrule(l){2-4} 
                                    & 39 & \texttt{flow\_total\_ocpp16\_starttransaction\_packets}         & The total number of the OCPP 1.6 StartTransaction messages                                                                 \\ \cmidrule(l){2-4} 
                                    & 40 & \texttt{flow\_total\_ocpp16\_remotestarttransaction\_packets}   & The total number of the OCPP 1.6 RemoteStartTransaction messages                                                           \\ \cmidrule(l){2-4} 
                                    & 41 & \texttt{flow\_total\_ocpp16\_authorize\_not\_accepted\_packets} & The total number of Authorize.conf messages containing an "Invalid", "Blocked" or "Expired" AuthorizationStatus            \\ \cmidrule(l){2-4} 
                                    & 42 & \texttt{flow\_total\_ocpp16\_setchargingprofile\_packets}       & The total number of the OCPP 1.6 SetChargingProfile messages                                                              \\ \cmidrule(l){2-4} 
                                    & 43 & \texttt{flow\_avg\_ocpp16\_setchargingprofile\_limit}           & Average number of the "limit" value of SetChargingProfile messages                                                         \\ \cmidrule(l){2-4} 
                                    & 44 & \texttt{flow\_max\_ocpp16\_setchargingprofile\_limit}           & Maximum value of the "limit" value of SetChargingProfile messages                                                          \\ \cmidrule(l){2-4} 
                                    & 45 & \texttt{flow\_min\_ocpp16\_setchargingprofile\_limit}           & Minimum value of the "limit" value of SetChargingProfile messages                                                          \\ \cmidrule(l){2-4} 
                                    & 46 & \texttt{flow\_avg\_ocpp16\_setchargingprofile\_minchargingrate} & Average number of the "minChargingRate" attribute of SetChargingProfile messages                                           \\ \cmidrule(l){2-4} 
                                    & 47 & \texttt{flow\_min\_ocpp16\_setchargingprofile\_minchargingrate} & Minimum value of the "minChargingRate" attribute of SetChargingProfile messages                                            \\ \cmidrule(l){2-4} 
                                    & 48 & \texttt{flow\_max\_ocpp16\_setchargingprofile\_minchargingrate} & Maximum value of the "minChargingRate" attribute of SetChargingProfile messages                                            \\ \cmidrule(l){2-4} 
                                    & 49 & \texttt{flow\_total\_ocpp16\_metervalues}                       & The total number of meterValues messages                                                                                   \\ \cmidrule(l){2-4} 
                                    & 50 & \texttt{flow\_min\_ocpp16\_metervalues\_soc}                    & The minimum value of State of Charge attribute of the meterValues messages                                                 \\ \cmidrule(l){2-4} 
                                    & 51 & \texttt{flow\_max\_ocpp16\_metervalues\_soc}                    & The maximum value of State of Charge attribute of the meterValues messages                                                 \\ \cmidrule(l){2-4} 
                                    & 52 & \texttt{flow\_avg\_ocpp16\_metervalues\_wh\_diff}               & The average of the difference between the "Energy.Active.Import.Register" attributes of consequentive meterValues messages \\ \cmidrule(l){2-4} 
                                    & 53 & \texttt{flow\_max\_ocpp16\_metervalues\_wh\_diff}               & The maximum difference between the "Energy.Active.Import.Register" attributes of consequentive meterValues messages        \\ \cmidrule(l){2-4} 
                                    & 54 & \texttt{flow\_min\_ocpp16\_metervalues\_wh\_diff}               & The minimum difference between the "Energy.Active.Import.Register" attributes of consequentive meterValues messages        \\ \midrule
\multicolumn{1}{c}{Other}         & 55 & \texttt{label}                                                    & String that describes the classification result of the flow. It can be normal, unlabelled, or denote a specific cyberattack.                                                                                                                           \\ \bottomrule
\end{tabular}}
\end{table}

\section{Results}
\label{sec:Experimental Results}

\subsection{Experiment Setup}
\label{sec:Experiment Setup}

\begin{figure}[htbp]
    \centering
    \includegraphics[width=0.6\textwidth,page=2]{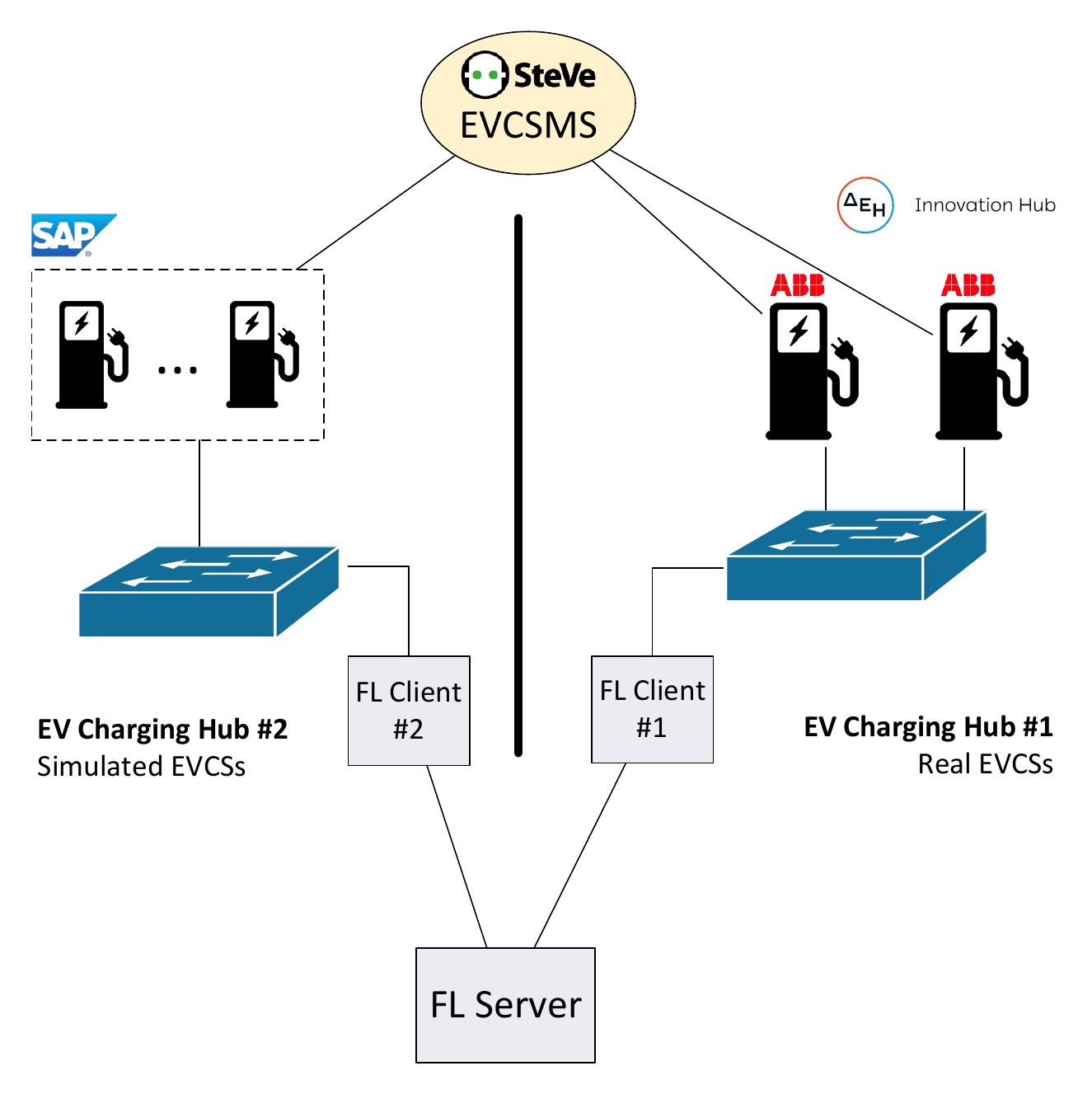}
    \caption{The Experimental Setup}
    \label{fig:fl-setup}
\end{figure}

\autoref{fig:fl-setup} depicts the experimental infrastructure utilized to test and evaluate the proposed solution. In this setup, we implemented the FL-based system model of section 2.1 by replicating two EV charging hubs. The first hub consists of real EV charging stations, namely a Terra Alternating Current (AC) 22kW wallbox type 2 (TAC-W22-T-0) and a Terra 54 Direct Current (DC) 50kW Fast Charger, both manufactured by ABB. The first hub is provided by the e-mobility laboratory of the Public Power Corporation (PPC) Innovation Hub\footnote{https://innovationhub.dei.gr/en/services/testing/other/e-mobility-laboratory/}. The second EV charging hub is composed of multiple virtual EV charging stations, which are simulated using the e-mobility charging stations simulator by SAP\footnote{https://github.com/sap/e-mobility-charging-stations-simulator}. Both hubs are managed by an EVCSMS, which is provided by the SteVe\footnote{https://github.com/steve-community/steve} open-source software. On both locations, the attacker utilises custom scripts written in Python in order to implement the cyberattacks described in section 2.2.

For implementing the FDI attacks, the \texttt{Ettercap}\footnote{https://www.ettercap-project.org/} tool is employed to conduct ARP poisoning. This procedure aims to poison the ARP cache of the EVCSs by giving the false information that the attacker's machine is the default gateway that the EVCSs need to route their packets to in order to reach the EVCSMS. Next, the appropriate \texttt{iptables} rules are inserted into the attacker's machine, in order to redirect incoming traffic from the EVCS to a NetFilter queue, allowing access and further manipulation of the packet via external software. Then, the \texttt{NetFilterQueue}\footnote{https://github.com/oremanj/python-netfilterqueue} library is utilized by the attacker in order to access the content of the NetFilter queue and manipulate the packets. For the manipulation process, the attacker utilises the \texttt{scapy} library. Finally, the packet is sent back to the network for its original destination.

For the flooding attacks, the attacker utilizes the \texttt{multiprocessing} package of Python to spawn multiple processes that act as separate EVCS bots. Then, each bot launches multiple processing threads, each thread representing an EVCS. Each thread uses the \texttt{websockets} Python library to initiate a WebSocket session with the target EVCSMS. If the connection fails, the thread tries again by randomly changing the EVCS ID. If the connection is accepted, the EVCS thread sends a \texttt{BootNotification.req} and then subsequent \texttt{Heartbeat.req} messages, each 1 second or more frequently.

\begin{table}[htbp]
\centering
\caption{Detection Capabilities and Relevant Features of CICFlowMeter and OCPPFlowMeter}
\label{tab:Association of cyberattacks with network flow modules and features}
\resizebox{\columnwidth}{!}{%
\begin{tabular}{lccc}
    \toprule
    \textbf{OCPP Cyberattack}                                       & \textbf{CICFlowMeter}        & \textbf{OCPPFlowMeter}        & \textbf{Relevant OCPPFlowMeter Features} \\\hline
    Charging Profile Manipulation     & {\color{DarkRed}\tikzxmark}  & {\color{DarkGreen}\checkmark} & \texttt{flow\_max\_ocpp16\_setchargingprofile\_limit} \\\hline
    Denial of Charge              & {\color{DarkRed}\tikzxmark}  & {\color{DarkGreen}\checkmark} &  \makecell[c]{\texttt{flow\_total\_ocpp16\_starttransaction\_packets},\\\texttt{flow\_total\_ocpp16\_authorize\_not\_accepted\_packets},\\\texttt{flow\_total\_ocpp16\_remotestarttransaction\_packets}} \\\hline
    Heartbeat Flood               & {\color{DarkGreen}\checkmark} & {\color{DarkGreen}\checkmark} & \makecell[c]{\texttt{src\_ip}, \texttt{dst\_ip}, \texttt{total\_flow\_packets},\\\texttt{total\_fw\_packets}, \texttt{total\_bw\_packets},  \texttt{flow\_total\_PSH\_flag},\\\texttt{flow\_total\_ACK\_flag}, \texttt{flow\_total\_websocket\_data\_messages},\\\texttt{flow\_total\_ocpp16\_heartbeat\_packets}} \\\hline
    EVCS Session Establishment Flood  & {\color{DarkGreen}\checkmark} & {\color{DarkGreen}\checkmark} & \makecell[c]{\texttt{flow\_total\_FIN\_flag}, \texttt{flow\_total\_http\_4xx\_packets},\\\texttt{flow\_total\_http\_get\_packets}}\\ 
    \bottomrule
\end{tabular}}
\end{table}

The evaluation results were calculated by using the data of both the \texttt{CICFlowMeter} and the \texttt{OCPPFlowMeter}. As discussed in section 2.3, the \texttt{OCPPFlowMeter} focuses on the OCPP features for generating network flows, enabling the detection of more attacks against OCPP. \autoref{tab:Association of cyberattacks with network flow modules and features} summarizes the capabilities of each network flow module with respect to the attacks implemented in the evaluation as well as the most prominent \texttt{OCPPFlowMeter} features for each attack. 

 Finally, the overall dataset consists of 4,415 samples, with each sample corresponding to a network flow. The dataset consists of four attack classes, as described above, and one normal, indicating benign traffic. Moreover, the dataset is balanced, meaning that all classes are represented by an equal number of samples. We consider that 3 clients participate in the FL, with the dataset evenly divided among them to ensure each client receives an equal number of labels. Each client then splits its local dataset into training, validation, and test sets, using a ratio of 0.6, 0.1, and 0.3, respectively. Regarding the FL training setting, the local model of each FL client is a neural network consisting of 3 fully connected hidden layers with 128 neurons and ReLU activation. The training consisted of 40 rounds with a batch size of 32 and learning rate equal to 0.001. 

\subsection{Detection Results}
\label{sec:Detection Results}

Before analyzing the detection performance of the proposed FL-based IDS, the relevant evaluation metrics are introduced first. On the one hand, True Positives (TP) denotes the number of correct classifications with respect to the presence of the attacks. Similarly, True Negatives (TN) indicates the number of correct classifications regarding the normal network flows. On the other hand, False Negatives (FN) and False Positives (FP) imply mistaken classifications related to the attacks. Therefore, based on the aforementioned terms, the following evaluation metrics are used:

\begin{equation}
Accuracy = \frac{TP+TN}{TP+TN+FP+FN}
\label{accuraccy}
\end{equation}

\begin{equation}
TPR = \frac{TP}{TP+FN}
\label{tpr}
\end{equation}

\begin{equation}
FPR = \frac{FP}{FP+TN}
\label{fpr}
\end{equation}

\begin{equation}
F1 = \frac{2 \times TP}{2\times TP+FP +FN} 
\label{f1}
\end{equation}

\autoref{tab:Evaluation Results of the Proposed FL Architecture With Various Aggregation Methods - CICFlowMeter} and \autoref{tab:Evaluation Results of the Proposed FL Architecture With Various Aggregation Methods - OCPPFlowMeter} summarize the evaluation results of the proposed FL-based IDS, for the two network flow extraction modules, by trying 6 different aggregation methods: (a) FedAvg, (b) FedProx, (c) FedAdam, (d) FedAdagrad, (e) FedYogi, and (f) FedTree. 

Based on the evaluation results for the \texttt{CICFlowMeter} module, FedProx achieves the best performance where $Accuracy=99.18\%$, $TPR=99.18\%$, $FPR=0.16\%$ and $F1=99.36\%$. On the contrary, the worst performance is calculated by FedAdagrad and FedTree where $Accuracy=98.26\%$, $TPR=98.26\%$, $FPR=0.21\%$ and $F1=99.18\%$.

For the \texttt{OCPPFlowMeter} module, FedAvg, FedProx and FedTree achieve the best performance where $Accuracy=99.21\%$, $TPR=99.21\%$, $FPR=0.20\%$ and $F1=99.21\%$. On the contrary, the worst performance is calculated by FedAdagrad where $Accuracy=79.78\%$, $TPR=79.78\%$, $FPR=5.05\%$ and $F1=72.87\%$.

 Finally, for both \texttt{CICFlowMeter} and \texttt{OCPPFlowMeter}, we evaluate the performance of the centralized method, which represents a scenario where all data are collected at a single node that performs the model training. Naturally, the centralized method serves as an upper performance bound. However, it is evident that FL achieves nearly the same performance as the centralized approach, while also inherently ensuring privacy, a capability the centralized method lacks.
\begin{table}[htbp]
\centering
\caption{Evaluation Results of the Proposed FL Architecture With Various Aggregation Methods - CICFlowMeter}
\label{tab:Evaluation Results of the Proposed FL Architecture With Various Aggregation Methods - CICFlowMeter}
\begin{tabular}{l| l l l l l} 
\hline 
\textbf{Strategy} & \textbf{Accuracy} & \textbf{TPR} & \textbf{FPR} & \textbf{F1}\\
\hline
FedAvg & 99.36\% & 99.36\% & 0.23\% & 99.08\% \\
\hline
FedProx & 99.18\% & 99.18\% & 0.16\% & 99.36\% \\
\hline
FedAdam & 99.18\% & 99.18\% & 0.21\% & 99.18\% \\
\hline
FedAdagrad & 98.26\% & 98.26\% & 0.21\% & 99.18\% \\
\hline
FedYogi & 99.45\% & 99.45\% & 0.44\% & 98.25\% \\
\hline
FedTree & 98.26\% & 98.26\% & 0.21\% & 99.18\% \\
\hline
Centralized & 99.52\% & 99.52\% & 0.15\% & 99.52\% \\
\hline
\end{tabular}
\end{table}

\begin{table}[htbp]
\centering
\caption{Evaluation Results of the Proposed FL Architecture With Various Aggregation Methods - OCPPFlowMeter}
\label{tab:Evaluation Results of the Proposed FL Architecture With Various Aggregation Methods - OCPPFlowMeter}
\begin{tabular}{l| l l l l l} 
\hline 
\textbf{Strategy} & \textbf{Accuracy} & \textbf{TPR} & \textbf{FPR} & \textbf{F1}\\
\hline
FedAvg & 99.21\% & 99.21\% & 0.20\% & 99.21\% \\
\hline
FedProx & 99.21\% & 99.21\% & 0.20\% & 99.21\% \\
\hline
FedAdam & 99.14\% & 99.14\% & 2.15\% & 99.14\% \\
\hline
FedAdagrad & 98.49\% & 98.49\% & 4.05\% & 98.37\% \\
\hline
FedYogi & 99.07\% & 99.07\% & 0.23\% & 98.07\% \\
\hline
FedTree & 99.21\% & 99.21\% & 0.20\% & 99.21\% \\
\hline
Centralized & 99.66\% & 99.66\% & 0.08\% & 99.66\% \\
\hline
\end{tabular}
\end{table}

\section{Discussion}

In this paper, an FL-based IDS is presented, which aims to detect cyberattacks against the EV charging infrastructure based on the OCPP 1.6. The proposed system realizes multiple FL clients on multiple EV charging hubs, which analyze the local OCPP 1.6 network traffic in terms of network flows and contribute to the training of a global AI model. Moreover, the FL client integrates the \texttt{OCPPFlowMeter}, a new tool for network flows that generates network flow statistics relevant to OCPP 1.6, thus assisting in the detection of both flooding and FDI attacks.

An experimental setup was described, based on both simulated and real EV charging stations, showcasing high detection performance. By comparing the results from 6 FL aggregation methods, it is concluded that the FredProx, FedAvg and FedTree provided better results, especially in terms of FPR and F1 score.

However, it should be noted that a cyberattack is detected only by assuming that the detector is able to capture the relevant malicious activity. If the attacker is able to avoid the packet capture, or if the attacker leverages adversarial AI techniques to evade the detection from the AI models, then the attack may remain undetectable. In these cases, an attack could be detected by observing the state of the system, i.e. the symptoms of a potential attack. Moreover, while OCPP 1.6 is considered dominant  in the market at the time of writing, future versions of OCPP (e.g., OCPP 2.0.1) may require the revision of the \texttt{OCPPFlowMeter} tool to ensure support.  In addition, the proposed FL architecture assumes that each EV charging hub is equipped with a host machine and network switch with port mirroring capabilities, thus requiring additional investments from the end user. Finally, concerning the computational requirements of training the AI models, the employed models do not result in a resource-intensive task, since they can be considered as lightweight. 

Considering the aforementioned remarks, as future work, we plan to extend our detection method by working on the following points: a) detecting an attack not only by its traces, but by assessing also the system status and performance Key Performance Indicators (KPIs) that would indicate the potential impact of a cyberattack, b) strengthening the resilience of our AI models against adversarial attacks, c) extending our threat analysis and the \texttt{OCPPFlowMeter} tool to OCPP 2.0.1. 

\section{Conclusion}

In the present work, we study the detection of cyberattacks against OCPP 1.6 by introducing an FL-based IDS. The literature review identified several IDS solutions. Some of the existing IDS adopt the FL paradigm while others focus on applying deep learning models that are trained in a centralized manner. Moreover, the solutions that detect threats against the EVCS infrastructure, mainly utilize centrally trained models, while their detection methodology rely on analyzing the network and transport layer attributes of network flows. Furthermore, the threat model considered by these solutions focus on generic network threats that are applicable to multiple IoT domains rather than application-specific threat scenarios.

Considering the aforementioned remarks, we developed a privacy-preserving FL-based anomaly detection method which relies on network flow features, generated by \texttt{OCPPFlowMeter}, that consider not only network and transport layer characteristics, but also features related to the underlying application protocols of WebSocket, HTTP and OCPP. To assess our methodology, we described four cyberattacks against OCPP, with two of them being detectable only by observing OCPP-specific features. The AI-based analysis of those features enable the detection of all four attacks, compared to the conventional analysis of \texttt{CICFlowMeter} features. Finally, a comparative analysis of six FL aggregation methods is presented, which are compared to the centralized training approach that serves as the upper performance bound, indicating that our FL-based method achieves nearly the same performance while retaining the benefit of data confidentiality.

\section*{Data Availability}
The data producing the results of this work are published on Zenodo\footnote{\url{https://zenodo.org/records/14887131}} and IEEE Dataport\footnote{\url{https://ieee-dataport.org/documents/federated-ocpp-16-intrusion-detection-dataset}}, titled as \textit{Federated OCPP 1.6 Intrusion Detection Dataset}.

\section*{Acknowledgments}
This project has received funding from the European Union’s Horizon 2020 and Horizon Europe research and innovation programmes under grant agreement No 101021936 (ELECTRON) and No 101070455 (DYNABIC). Disclaimer: Funded by the European Union. Views and opinions expressed are, however, those of the author(s) only and do not necessarily reflect those of the European Union or European Commission. Neither the European Union nor the European Commission can be held responsible for them.

\bibliographystyle{unsrtnat}
\bibliography{refs}

\end{document}